# Nano-scale characterization of the formation of silver layers during electroless deposition on polymeric surfaces

Aniruddha Dutta[1, 2, *], Biao Yuan[5], Helge Heinrich[1, 2], Christopher N. Grabill[3], Stephen M. Kuebler[3, 4], Aniket Bhattacharya[1]

[1]Department of Physics; [2]Advanced Materials Processing and Analysis Center;
[3]Department of Chemistry; [4]CREOL, The College of Optics and Photonics; and
[5]Department of Mechanical, Materials and Aerospace Engineering
University of Central Florida, Orlando, FL 32816.

*aniruddha@knights.ucf.edu

## Abstract

*We report here a quantitative method of Transmission Electron Microscopy (TEM) to measure the shapes, sizes and volumes of nanoparticles which are responsible for their properties. Gold nanoparticles (Au NPs) acting as nucleating agents for the electroless deposition of silver NPs on SU-8 polymers were analyzed in this project. The atomic-number contrast (Z-contrast) imaging technique reveals the height and effective diameter of each Au NP and a volume distribution is obtained. Varying the reducing agents produced Au NPs of different sizes which were found both on the polymer surface and in some cases buried several nanometers below the surface. The morphology of Au NPs is an important factor for systems that use surface-bound nanoparticles as nucleation sites as in electroless metallization. Electrolessly deposited silver layers reduced by hydroquinone on SU-8 polymer are analyzed in this project.*

**Keywords**: Polymer, gold, silver, HAADF-STEM.

## 1. Introduction

Determining the sizes and shapes of metal NPs is crucial in understanding the morphology associated with nonmaterial systems. Various reducing agents effect the formation of Au NPs at the surface of SU-8 cross-linked polymers which are increasingly employed for micro and nano-scale fabrication [1, 2]. In this work Au NPs have been formed on the SU-8 polymer surface by in-situ reduction. Quantitative TEM has been used for the evaluation of NP sizes, volumes, structures, and compositions, and to understand their distribution on the surface [3-7].

## 2. Experimental

Gold and silver NPs deposited on SU-8 polymer samples were provided by the group of Dr. Stephen Kuebler of the Chemistry Department at UCF. Au NPs were formed by binding Au cations to an aminated polymer surface immersed in an aqueous solution of $5.3 \times 10^{-4}$ M $HAuCl_4$. The solution was then reduced with 0.1M $NaBH_4$. All depositions were done in ambient temperature. Preparation of the Ag NPs was done according to the method introduced by Khalid *et al.* [8]; the reduction by citrate was accomplished by immersing samples for eight hours in aqueous 1% (w/v) sodium citrate. Reduction with hydroquinone was achieved by immersing samples for one hour in aqueous 0.1 M hydroquinone.

Samples for plan-view TEM imaging were prepared by scraping a thin section from the surface of the polymer film. Cross sectional TEM samples were prepared by Focused Ion Beam (FIB) thinning. Each cross-sectional sample was coated with 30 to 50 nm carbon followed by 10 nm Au-Pd coating before the FIB cutting to increase the contrast for TEM micrographs. A Tecnai F30 TEM at 300 kV was used for plan-view and cross-sectional imaging.

### 3. Results and discussion

*3.1. Characterization of Au NPs formed by $NaBH_4$ and hydroquinone reduction.*

Fig1 (a) shows a plan-view TEM image of Au NPs deposited on the surface of a SU-8 polymer after reduction by $NaBH_4$. The nanoparticles appear evenly spaced with no long-range ordering. Energy-Dispersive X-ray Spectroscopy(EDS) was performed on these small 2 to 5 nm diameter particles confirming the nanoparticles consisted of gold without any Pd. Electron diffraction proved that the particles have a face centered cubic(FCC) structure. Fig 1(b) represents a High-Angle Annular Dark-Field (HAADF) Scanning Transmission Electron Microscopy (STEM) plan-view image confirming the homogeneity of the nanoparticle distribution.

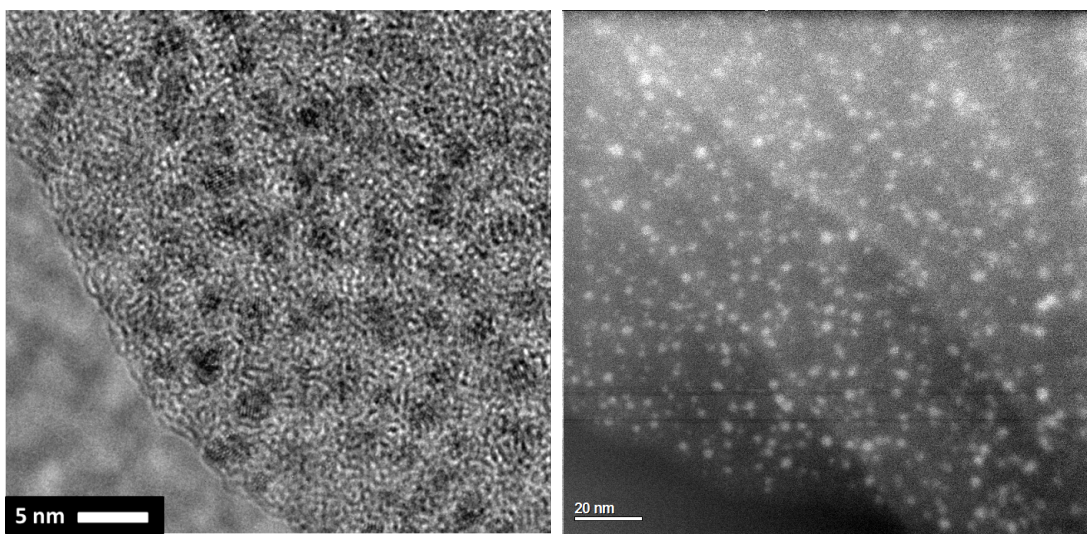

Fig1 (a). Bright-field plan-view TEM micrograph of Au NPs generated on the polymer by reduction of gold ions bound at the surface by ethylenediamine (ED) then reduced with $NaBH_4$. (b) HAADF-STEM plan view micrograph of $NaBH_4$-generated Au NPs.

The volumes, heights and the cross-sectional areas of individual NPs were obtained from intensity-calibrated HAADF-STEM micrographs. For a statistical analysis 150 NPs were considered. The particle size analysis reveals that the average volume of the Au NPs is 8 $nm^3$ with a Relative Standard Deviation (RSD) of the volume distribution of 56%. The NPs are typically not spherical but rather oblate or prolate spheroids. Fig 2(a) shows the volume distribution of the Au NPs for a plan-view sample. Most of the NPs have volume between 5-7 nm$^3$ with around 23% NPs being very small with volumes below 5 nm$^3$.

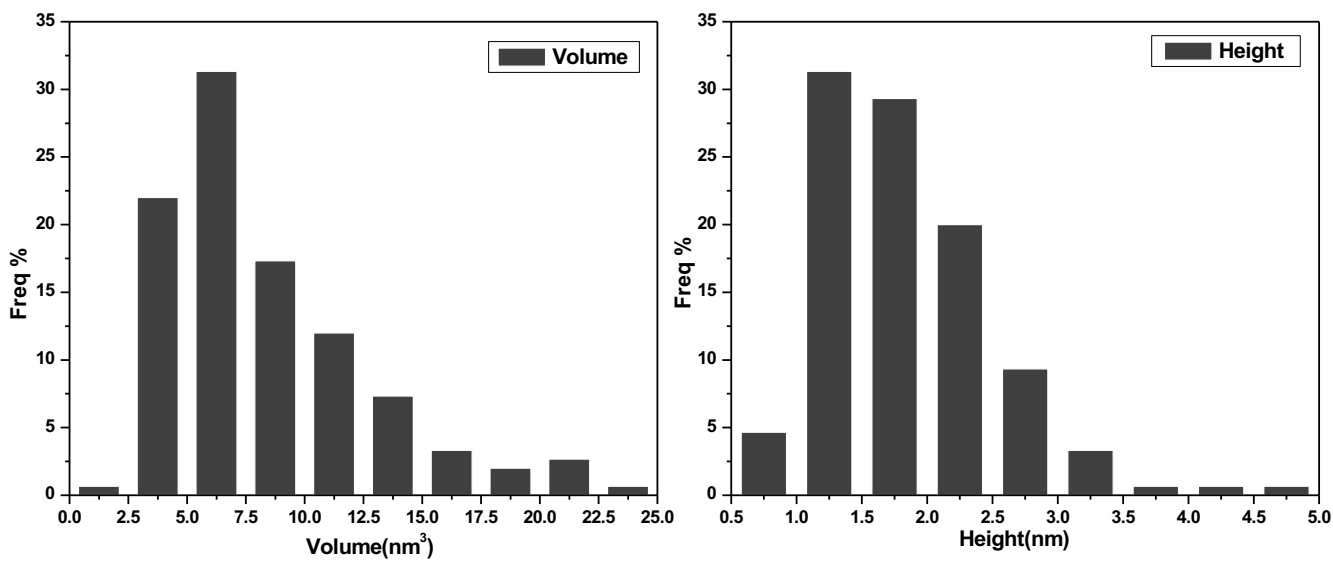

Fig 2(a). Volume distribution of 150 Au NPs. (b) Height distribution of Au NPs from calibrated HAADF-STEM micrographs.

Likewise, the measured particle heights from integrated pixel intensities results in an average particle height of 1.8 nm and a RSD of 37%. The mean lateral radius was calculated to be 1.4 nm with a RSD of 13%. From the height distribution in Fig 2(b) we infer that almost half of the particles are less than 2 nm in height.

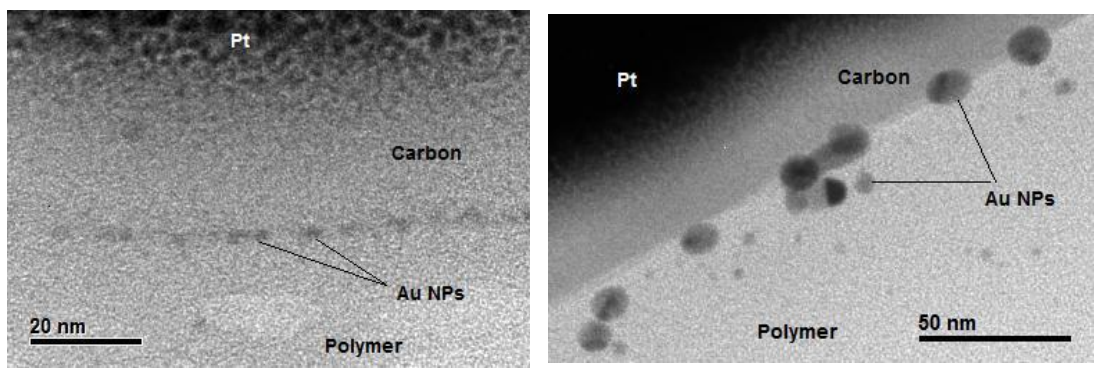

Fig 3(a) Cross-sectional Bright Field (BF) TEM micrographs of Au NPs reduced by NaBH$_4$. (b)Hydroquinone (HQ) reduced Au NPs on the SU-8 polymer surface.

Fig 3(a) shows the BF image of Au NPs with 2-5 nm in diameter reduced by NaBH$_4$ on the polymer surface. The Au NPs of 10 to20 nm in diameter in Fig. 3(b) are consistent with Au NPs reduced by HQ as reported in the literature [9]. Being a stronger reducing agent, with NaBH$_4$ much smaller Au NPs are created compared to HQ which has a positive reduction potential compared to NaBH$_4$[10-12].

## 3.2. Characterization of Ag NPs formed by reduction of hydroquinone from silver citrate.

Gold NPs are bound to the surface of SU-8 polymer by coordinating $AuCl_4^-$ at surface bound amine sites. Surface bound $Au^{3+}$ is then reduced with $NaBH_4$, causing Au NPs to form. Ag is deposited onto Au NPs via reduction of $Ag^+$ by hydroquinone for 6 hours in presence of Gum Arabic. Fig 4 shows the chemical reaction of formation of Ag NPs by the above method [1, 13].

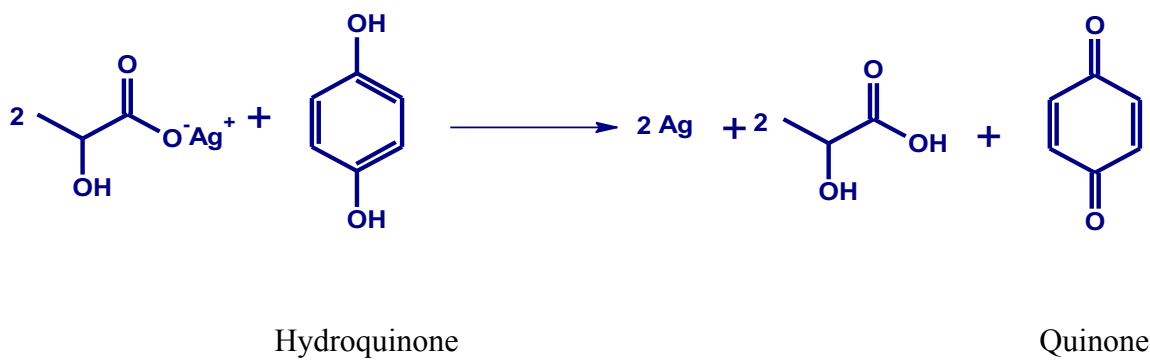

Fig 4. Reaction scheme for electroless silver deposition.

Fig 5 shows a plan-view HAADF-STEM micrograph of Ag NPs formed on SU-8 polymer. The Au NPs act as nucleation sites for the silver grains. Comparison of the plan-view images in Fig. 1(b) and Fig. 5 shows that the distance of small Au NPs is smaller than the distance of centers of Ag grains. We can therefore conclude that not all Au NPs are nucleation sites for large Ag grains. The average thickness of the silver layer was found to be 45 nm with a standard deviation of 18 nm. The granular structure of silver grains accounts for the large standard deviation. Gum Arabic, a complex protein (stabilizer), has been used in this case in the solution to control the silver deposition. The average volume for the Ag grains was $(65000 \pm 9000)$ $nm^3$ with an average diameter as 45 nm.

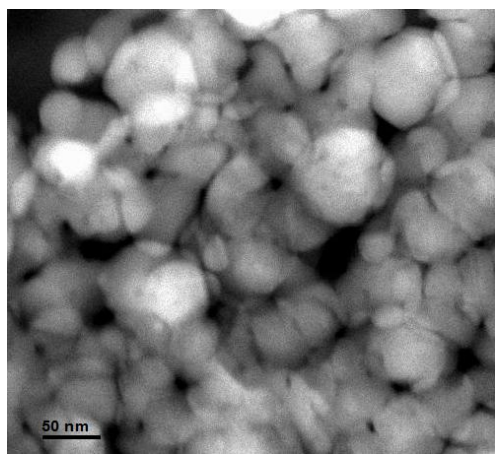

Fig 5. Plan-view HAADF-STEM of Ag NPs on a SU8 polymer prepared by 6 hours of electroless deposition using $[Ag+] = 5.6$ mM in the presence of gum Arabic [15].

## 4. Conclusions

The above work suggests that the size of Au NPs generated can be controlled with choice of reducing agents. NPs that are generated on the surface by HQ are much larger than those generated beneath the surface. This happens probably because the rapid reaction of $NaBH_4$ with Au ions on the surface promotes diffusion resulting in growth rather than nucleation. The model also suggests that NPs can be prepared at different depths below the surface for several reducing agents. These NPs are useful for creating material systems with controlled optical enhancement that are not compromised by fluorescence quenching. It follows then that the role of the polymer matrix may be important in determining, or even controlling, the shape, size, and properties of the NPs generated. Controlling the rate of silver growth and shaping its 3D morphology will be the area of further studies.

**Acknowledgements**

This work was supported by NSF CAREER grant #0748712 and NSF-CHE grant #0809821.We acknowledge Dr. Aniket Bhattacharya and his group from University of Central Florida (UCF) Physics Department for modeling and simulations. We also thank all the staff members of Advanced Material Processing & Analysis Center (AMPAC) at UCF for providing support and help with handling of the instruments.